# Giant enhancement in critical current density, up to a hundredfold, in superconducting NaFe$_{0.97}$Co$_{0.03}$As single crystals under hydrostatic pressure

Babar Shabbir[1], Xiaolin Wang[1*], S. R. Ghorbani[1,2], A. F. Wang[3], Shixue Dou[1], and X.H. Chen[3]

[1]Spintronic and Electronic Materials Group, Institute for Superconducting and Electronic Materials, Faculty of Engineering, Australian Institute for Innovative Materials, University of Wollongong, North Wollongong, NSW 2522, Australia

[2]Department of Physics, Ferdowsi University of Mashhad, Mashhad 9177948974, Iran

[3]Department of Physics, University of Science and Technology of China, 230026 Hefei, P. R. China

**Tremendous efforts towards improvement in the critical current density "$J_c$" of iron based superconductors (FeSCs), especially at relatively low temperatures and magnetic fields, have been made so far through different methods, resulting in real progress. $J_c$ at high temperatures in high fields still needs to be further improved, however, in order to meet the requirements of practical applications. Here, we demonstrate a simple approach to achieve this. Hydrostatic pressure can significantly enhance $J_c$ in NaFe$_{0.97}$Co$_{0.03}$As single crystals by at least tenfold at low field and more than a hundredfold at high fields. Significant enhancement in the in-field performance of NaFe$_{0.97}$Co$_{0.03}$As single crystal in terms of pinning force density ($F_p$) is found at high pressures. At high fields, the $F_p$ is over 20 and 80 times higher than under ambient pressure at 12K and 14K, respectively, at $P$=1GPa. We believe that the Co-doped NaFeAs compounds are very exciting and deserve to be more intensively investigated. Finally, it is worthwhile to say that by using hydrostatic pressure, we can achieve more milestones in terms of high $J_c$ values in different superconductors.**

Significant advances towards high performing Fe-based superconducting materials (FeSCs) have been made so far, because the combination of reasonable value of critical temperature ($T_c$), extremely high upper critical field ($H_{c2}$) on the order of 100T, high intrinsic pinning potential, low anisotropy (generally between 1-8), and high irreversible field ($H_{irr}$) makes this class of superconductors particularly attractive for large current and high field applications, where the critical current density ($J_c$) is a major limiting factor [1-14]. Therefore, improvement in $J_c$ by using various methods has also been one of the most important topics in the superconductivity research field. Texturing procedures, ion implantation/irradiation, and chemical doping are the most common approaches to enhancing $J_c$ in different superconductors. Although $J_c$ values can be improved by these methods, the major drawbacks are that $J_c$ decays rapidly in high fields, especially at high temperatures. Therefore, the $J_c$ values of the Fe-based superconductors at high fields and temperatures need to be improved. Furthermore, $T_c$ and low field $J_c$ deteriorate significantly for various types of superconductors under these approaches, which make them rather impractical for application. The reported enhancement in $J_c$ values at high fields and temperatures is still not more than one order of magnitude. [5, 15-22]. Generally, the requirements for enhancing $J_c$ in superconductors include $T_c$ enhancement, which can increase the effective superconducting volume, and the formation of more effective point pinning centres (related to the pinning mechanism).

Hydrostatic pressure has many significant impacts on Fe-based superconductors. For instance, pressure can raise the onset $T_c$ up to 50K at 1.5Gpa for $LaFeAsO_\delta$ [24]. The application of pressure on $BaFe_{1.92}Co_{0.08}As_2$ results in a very strong enhancement of $T_c$ from 11 to 21 K at 2.5 Gpa [25]. For Co-doped NaFeAs, enhancement in $T_c$ is more than 14K at 2.5GPa due to optimization of the structural parameters of the FeAs layers [26]. The $T_c$ of FeSe is enhanced up to 37K at 7Gpa [27]. Furthermore, pressure can induce reduction in anisotropy, more effective point pinning centres, and enhancement in $T_c$. We already anticipated in our previous case study that the most significant approach to enhancing Jc, particularly at high fields and temperatures, without degradation of Tc, is the use of hydrostatic pressure [23].

Most recent research regarding $J_c$ enhancement and pinning mechanisms is mainly focused on the 1111 system (RFeAsO, where R is a rare earth element), the 122 system ($BaFe_2As_2$, $Ba_{0.5}K_{0.5}Fe_2As_2$) and the iron chalcogenide 11 system. Only one report has revealed the nature of the pinning mechanism in LiFeAs (111 type FeSCs) so far, despite its simple structure and reasonable $T_c$ value as compared to the 1111 and 122 types [28]. NaFeAs ($T_c$ ≈10K) experiences three successive phase transitions around 52, 41, and 23 K, which can be related to structural, magnetic, and superconducting transitions, respectively [29-31]. Bulk superconductivity in NaFeAs with $T_c$ of ~20K can be achieved by the substitution of Co on Fe sites, which can suppress both magnetism and structural distortion [26, 32].The $T_c$ of $NaFe_{0.97}Co_{0.03}As$ single crystal is more sensitive to hydrostatic pressure as compared to other 11 and 111 Fe-based superconductors, and it has a large positive pressure coefficient [26]. In addition, $J_c$ values for Co-doped NaFeAs compounds have not been reported so far. Therefore, it is very interesting to see if the hydrostatic pressure can significantly improve the flux pinning for such compounds. High-quality NaFe$_{0.97}$Co$_{0.03}$As single crystals were grown by the conventional high temperature solution growth method using the NaAs self-flux technique [30]. In this communication, we report that hydrostatic pressure can enhance the $J_c$ by more than 100 times at high fields at 12 K and 14K in NaFe$_{0.97}$Co$_{0.03}$As single crystal. This is a giant enhancement of $J_c$ and a record high to the best of our knowledge. The $H_{irr}$ is improved by roughly 6 times at 14K under $P$=1GPa.

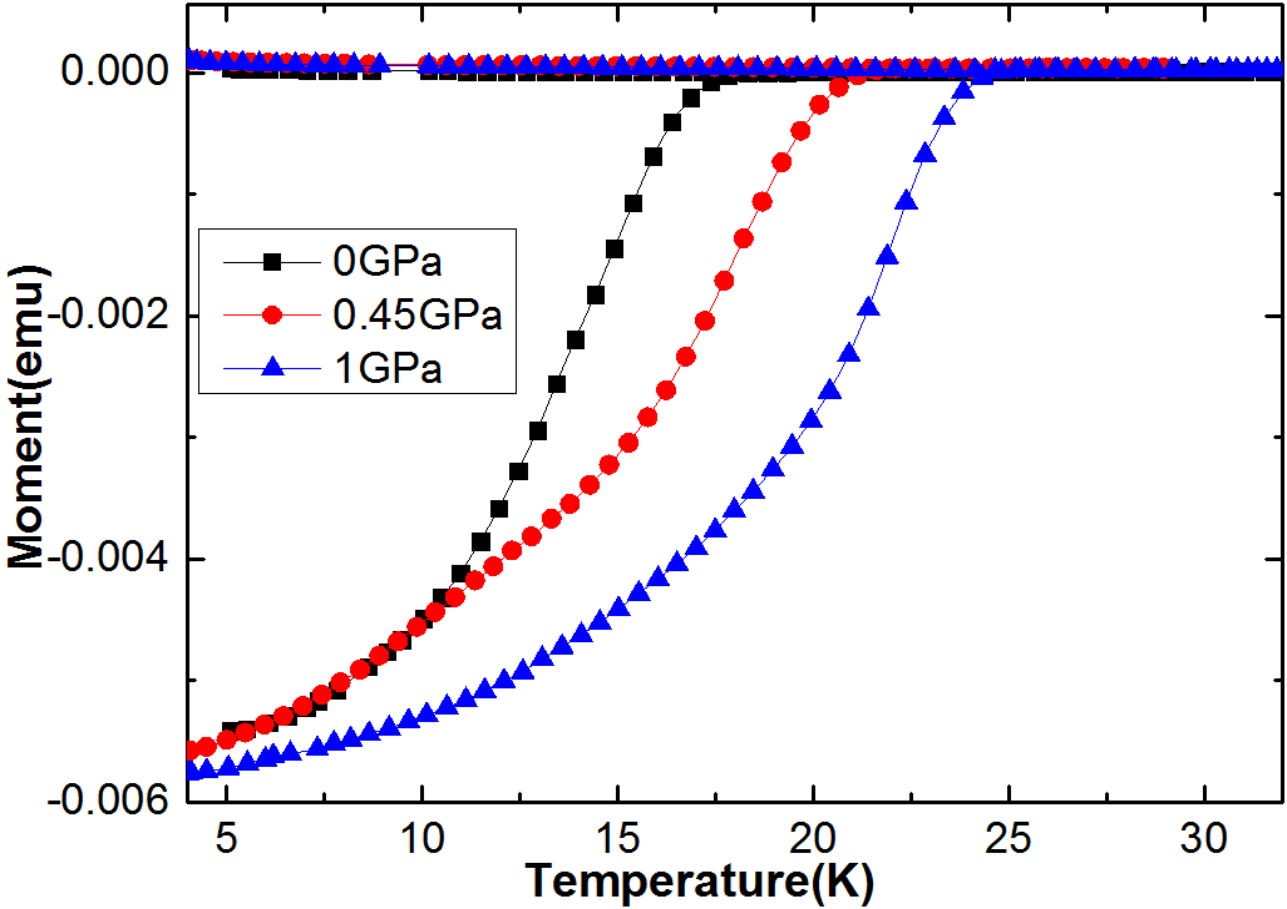

Figure 1: Temperature dependence of magnetic moment at different applied pressures in both ZFC and FC runs for NaFe0.97Co0.03As.

The temperature dependence of the magnetic moment for zero-field-cooled (ZFC) and field-cooled (FC) curves at different pressures are shown in Fig.1. The $T_c$ increases with pressure, from 17.95K for $P$ = 0 GPa to 24.33K for $P$ = 1 GPa, with a huge pressure coefficient, i.e. $dT_c/dP$~ 6.36K/GPa, which is nearly same to what we have already reported for NaFe$_{0.97}$Co$_{0.03}$As single crystal i.e. $dT_c/dP$= 7.06K·GPa$^{-1}$[26]. Interestingly, this pressure coefficient is more than two times greater than that of FeSe (3.2K·GPa$^{-1}$) [33]. The pressure-induced enhancement of $T_c$ in NaFe$_{0.97}$Co$_{0.03}$As can be associated with the optimization of the structural parameters of the FeAs layers, including the As–Fe–As bond angle and anion height [26, 34].

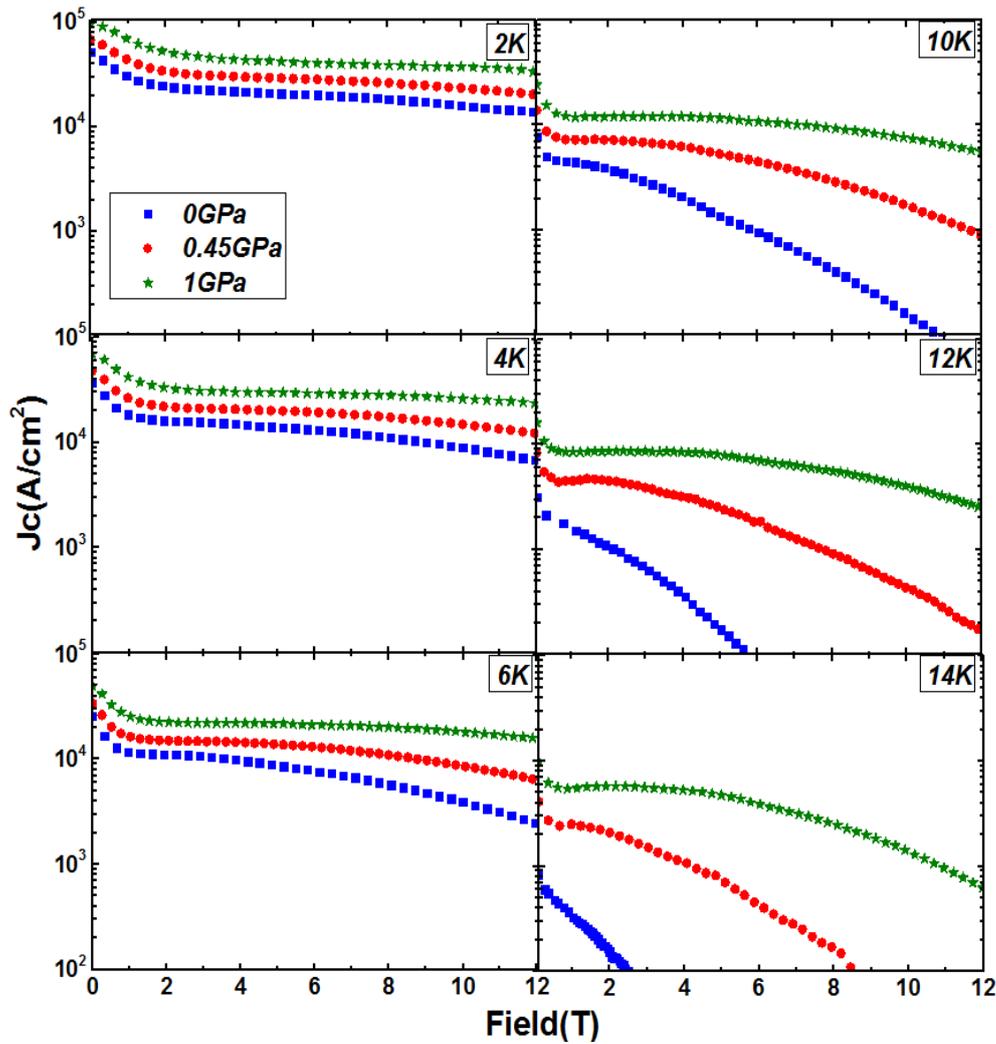

Figure 2. Field dependence of Jc at different pressures (0, 0.45 and 1GPa) at different temperatures.

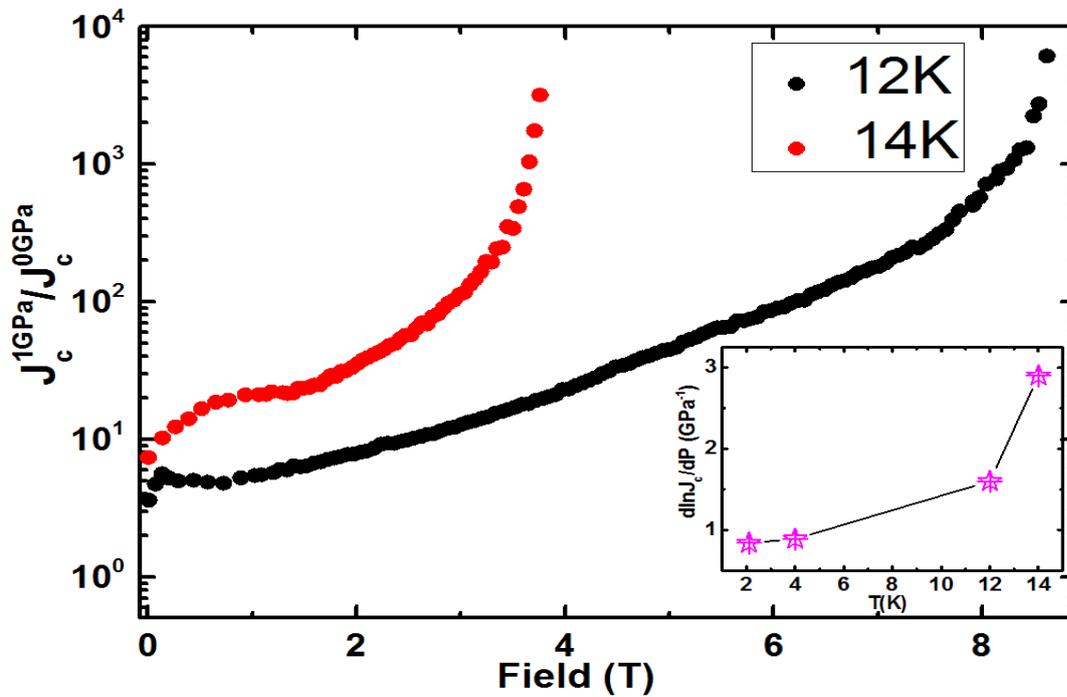

Figure 3: Plot of $J_c^{1GPa}/J_c^{0GPa}$ versus field at 12K and 14K. There is a giant enhancement in Jc values at P =1GPa. The inset shows the plot of d(lnJc)/dP versus temperature, which demonstrates enhancement in lnJc at a rate of nearly 3GPa-1 at 14K at The field dependence of $J_c$ at different temperatures obtained from the *M-H* curves by using Bean's model, at *P*=0GPa, *P*=0.45GPa and *P*=1GPa are shown in Fig.2. Remarkably, $J_c$ is increased significantly at both low and high fields, especially with enhancement of more than 10 times and up to more than 100 times for low and high fields at both 12 and 14 K, respectively. The significant positive effect of hydrostatic pressure on the $J_c$ at high fields and temperatures is further reflected in Fig.3, which shows the $J_c$ enhancement ratio (i.e. $J_c^{1GPa}/J_c^{0GPa}$) at 12 and 14K over a wide range of fields. We have taken the $J_c$ value at *P*=0GPa as a reference. The $J_c$ ratio values at both temperatures show significant improvements at low and high fields. Although this result also suggests that hydrostatic pressure is more effective at high fields and temperatures, it is worth mentioning that $J_c$ values are well improved at zero field at a significant rate, i.e. $d(\ln J_c)/dP$ = 1.6 and 2.9 GPa$^{-1}$ at 12 and 14K, respectively, as can be seen from the inset of Fig. 3. The $d(\ln J_c)/dP$ values that have been found are more significant than for yttrium barium copper oxide(YBCO) [35].

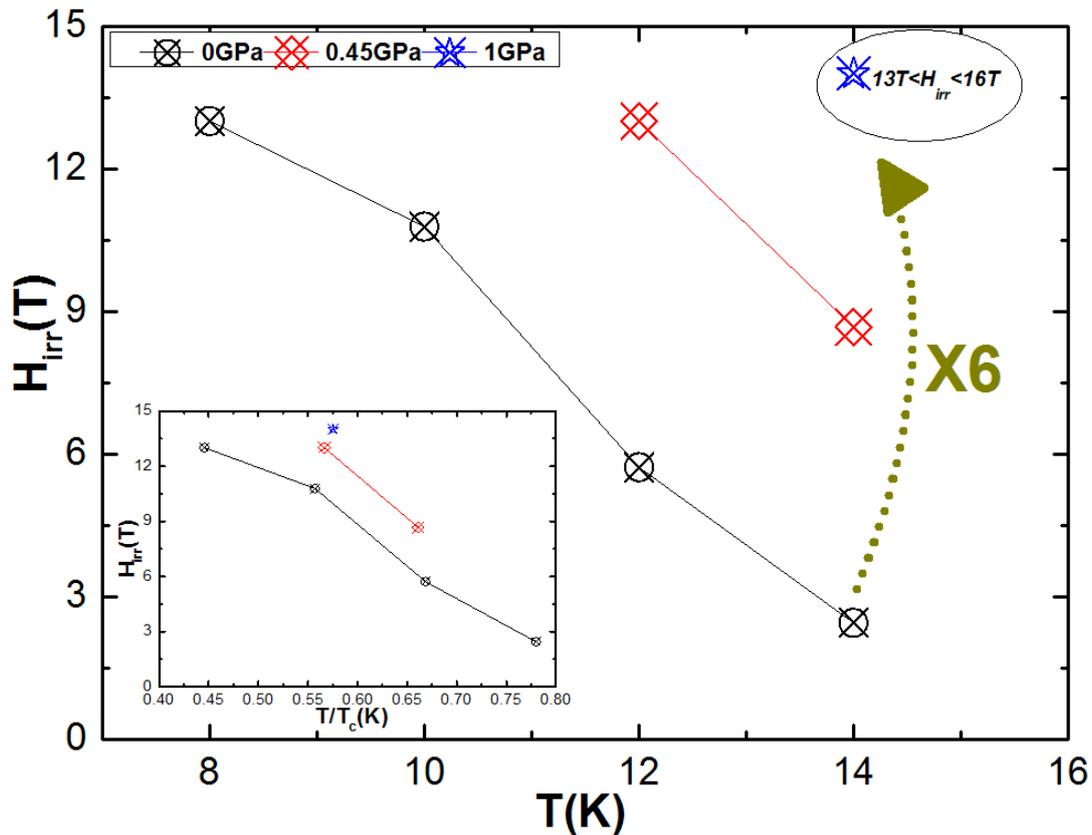

Figure 4: Plot of Hirr versus temperature at different pressures. Inset shows Hirr as a function of reduced temperature.

We also found that $H_{irr}$ of NaFe$_{0.97}$Co$_{0.03}$As is significantly increased by pressure. As shown in Fig.4, the $H_{irr}$ values improve gradually with pressure, and the H$_{irr}$ value at 14K is increased from 2.6 T at P=0 GPa up to 8.67T at P=0.45GPa and roughly more than 13 T at P=1GPa (by nearly six times).

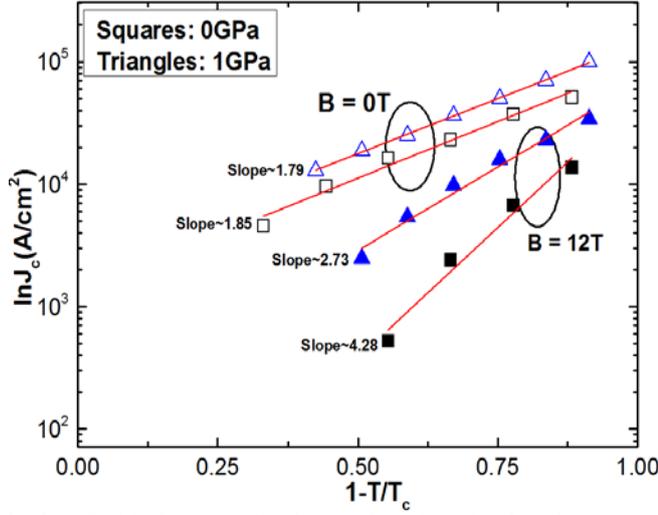

Figure 5: Logarithmic plot of critical current density as a function of reduced temperature at different pressures and magnetic fields.

In Fig.5, we show the temperature dependence of $J_c$ at 0 and 12T under different pressures. It follows power law [$J_c \propto (1-T/T_c)^\beta$] behaviour at different pressures. According to the Ginzburg-Landau theory, the exponent $\beta$ is used to identify different vortex pinning mechanisms at specified fields. It was found that $\beta = 1$ refers to non-interacting vortices and $\beta \geq 1.5$ corresponds to the core pinning mechanism [27]. The exponent $\beta$ (i.e. slope of the fitting line) is found to be 1.79 and 1.85 for zero field, and 2.73 and 4.28 at 12T, at 0 and 1GPa, respectively, which shows strong $J_c$ dependence on pressure. The low values of $\beta$ at $P=$1GPa indicate that the $J_c$ decays rather slowly in comparison to its values at $P=$0GPa. In addition, the differences between $P=$0GPa and $P=$1GPa scaling show a real pressure effect, the factor is roughly 2, which corresponds nicely to the low-T data in inset Fig. 3.

For polycrystalline bulks, high pressure can modify grain boundaries through reduction of the tunnelling barrier width and the tunnelling barrier height. The Wentzel-Kramers-Brillouin (WKB) approximation applied to a potential barrier gives the following simple expression [36-38]:

$$J_c = J_{co} \exp(-2kW) \quad (1)$$

Where $W$ is the barrier width, $k=(2mL)^{1/2}/\hbar$ is the decay constant, which depends on the barrier height $L$, $\hbar$ is the Planck constant, and $J_{c0}$ is the critical current density at 0K and 0T. The relative pressure dependence of $J_c$ can be obtained from Eq. (1) as [39]:

$$\frac{d \ln J_c}{dP} = \frac{d \ln J_{c0}}{dP} - \left[\left(\frac{d \ln W}{dP}\right) \ln\left(\frac{J_{c0}}{J_c}\right)\right] - \frac{1}{2}\left[\left(\frac{d \ln L}{dP}\right) \ln\left(\frac{J_{c0}}{J_c}\right)\right]$$

$$= \frac{d\ln J_{c0}}{dP} + \kappa_{GB} \ln\left(\frac{J_{c0}}{J_c}\right) + \frac{1}{2}\kappa_L \ln\left(\frac{J_{c0}}{J_c}\right) \quad (2)$$

Where the compressibility in the width and height of the grain boundary are defined by $\kappa_{GB} = -d\ln W/dP$ and $\kappa_L = -d\ln L/dP$, respectively. For single crystals, we assume to a first approximation that $\kappa_{GB}$ and $\kappa_L$ are roughly comparable, respectively, to the average linear compressibility values $\kappa_a = -d\ln a/dP$ ($\kappa_a \approx -0.029$ GPa$^{-1}$) and $\kappa_c = -d\ln c/dP$ ($\kappa_c \approx -0.065$ GPa$^{-1}$) of NaFe$_{0.97}$Co$_{0.03}$As crystal in the FeAs plane, where $a$ and $c$ are the in-plane and out-of-plane lattice parameters, respectively [27]. Therefore, we can write Eq. (2) as

$$\frac{d\ln J_c}{dP} \sim \frac{d\ln J_{c0}}{dP} + \kappa_a \ln\left(\frac{J_{c0}}{J_c}\right) + \frac{1}{2}\kappa_c \ln\left(\frac{J_{c0}}{J_c}\right) \quad (3)$$

By using $J_c \cong 1.3 \times 10^3$ A/cm$^2$ at 14 K and $J_{c0} \sim 10^5$ A/cm$^2$, we find that $(\kappa_a \ln(J_{c0}/J_c)) \approx 0.12$ GPa$^{-1}$ and $(1/2\kappa_c \ln(J_{c0}/J_c)) \approx 0.14$ GPa$^{-1}$, so that both of them together only contribute less than 10% to the already mentioned experimental value $d\ln J_c/dP$ = 2.09 GPa$^{-1}$ inset of Fig. 2b). This result suggests that the origin of the significant increase in $J_c(T)$ under pressure does not arise from the reduction of volume but mainly due to the pressure induced pinning centre phenomenon.

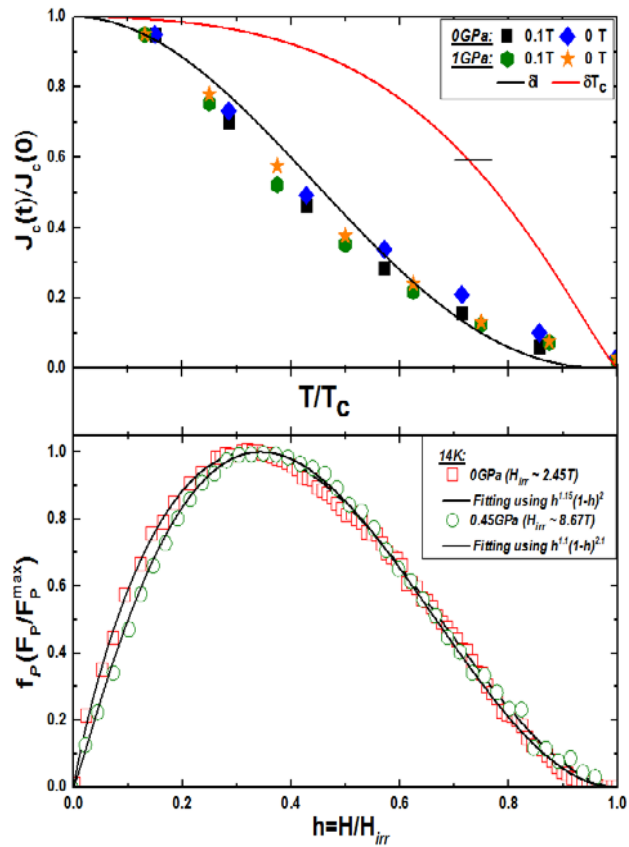

Figure 6: Top panel shows normalized temperature dependence ($t = T/T_c$) of normalized measured $J_c$ at 0.1T and 0T, in good agreement with $\delta\ell$ pinning. Lower panel shows plots of $F_p$ vs $H/H_{irr}$ at $P$=0GPa and $P$= 0.45GPa at 14K. The experimental data is fitted through the Dew-Hughes model, and the parameters are given in inset.

To gain further insight into the pressure effect on the pinning mechanism in NaFe$_{0.97}$Co$_{0.03}$As, the experimental results have been analysed by using collective pinning theory. There are two predominant mechanisms of core pinning, i.e. $\delta \ell$ pinning, which comes from spatial variation in the charge carrier mean free path, $\ell$, and $\delta T_c$ pinning due to randomly distributed spatial variation in $T_c$. According to the theoretical approach proposed by Griessen et al. [40], $J_c(t)/J_c(o) \propto (1-t^2)^{5/2}(1+t^2)^{-1/2}$ in case of $\delta \ell$ pinning, whereas $J_c(t)/J_c(o) \propto (1-t^2)^{7/6}(1+t^2)^{5/6}$ corresponds to $\delta T_c$ pinning, where $t = T/T_c$. Fig.6 (Top Panel) shows a comparison between the experimental $J_c$ values and the theoretically expected variation within the $\delta \ell$ and $\delta T_c$ pinning mechanisms at 0.1T and 0 T (the so-called remanent state shown by the solid symbols). The $J_c(t)$ values have been obtained from the $J_c(B)$ curves at various temperatures. It is found that the experimental data at $P=0$ and $P=1$GPa are in good agreement with theoretical $\delta \ell$ pinning. It is more likely that pinning in NaFe$_{0.97}$Co$_{0.03}$As originates from spatial variation of the mean free path "$\ell$". We observed similar results in BaFe$_{1.9}$Ni$_{0.1}$As$_2$ and SiCl$_4$ doped MgB$_2$ at low fields. In addition, $\delta \ell$ pinning has also been reported in FeTe$_{0.7}$Se$_{0.3}$ crystal [41-43]. In order to understand the nature of the pinning mechanisms in more detail, it is useful to study the variation of the vortex pinning force density, $F_p = J_c \times B$ with the field. The normalized pinning force density ($F_p/F_p^{max}$) as a function of reduced field ($H/H_{irr}$) at $P=0$GPa and $P=0.45$GPa at 14K is plotted in lower panel of Fig.6. $H_{irr}$ is estimated by using the criterion of $J_c \approx$ 100A/cm$^2$. We can use the Dew-Hughes formula, i.e. $F_p \propto h^m(1-h^n)$ to fit our experimental data, where $m$ and $n$ are fitting parameters to describe the nature of the pinning mechanism. We found that $m=1.15$ and $n=2$ at 0GPa, and $m=1.1$ and $n=2.1$ at 0.45GPa. According to the Dew-Hughes model, in the case of $\delta \ell$ pinning for a system dominated just by point pinning, the values of the fitting parameters are $m=1$ and $n=2$, with the $F_p^{normalized}$ maxima maximum occurring at $h_{max} = 0.33$, while $h_{max}$ occurs at 0.20 for surface/grain boundary pinning with $m=0.5$ and $n=2$. In case of $\delta T_c$ pinning, $h_{max}$ shifts to higher values, and the fitting parameters change accordingly. Further details can be found elsewhere [44]. The values of $m$ and $n$ that were found in the present study are almost the same at 0GPa and 0.45GPa, so normal core point pinning is dominant in our material.

Pressure can enhance the pinning force strength by a significant amount in NaFe$_{0.97}$Co$_{0.03}$As single crystal. The pinning force density as a function of field at 12K and 14K is plotted in Fig.7. At high field and pressures, the $F_p$ is found to be over 20 and 80 times higher than at 0 GPa at 12 and 14K, respectively.

Furthermore, pressure induces more point pinning centres at 12 K and 14 K, especially at P=1GPa, as can be seen in the inset of Fig.7.

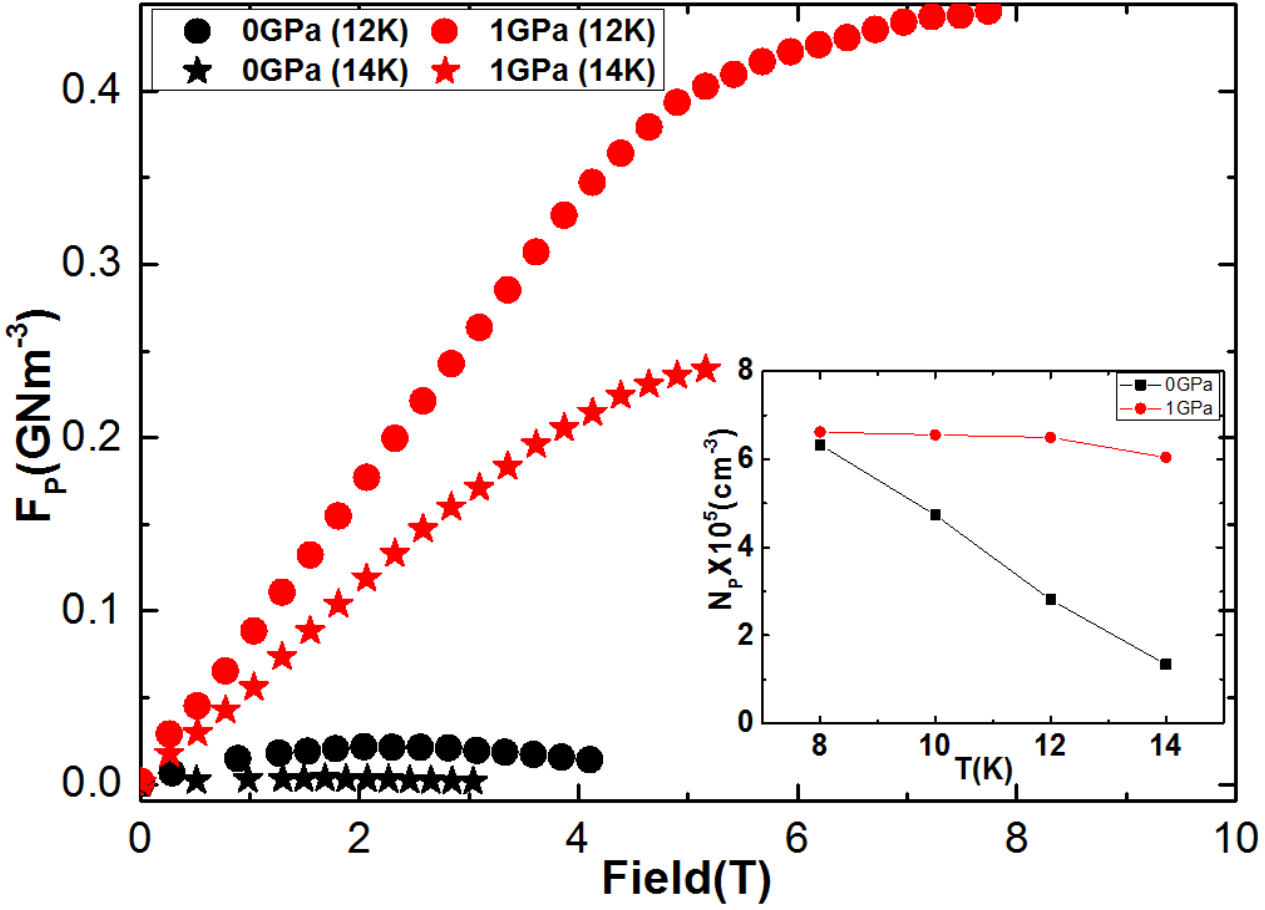

Figure 7: Pinning force density ($F_p$) as a function of field at P=0GPa and P=1GPa at 12K and 14K. The inset shows the temperature dependence of the pinning centre number density at the different pressures.

The number density of randomly distributed effective pinning centres ($N_p$) can be calculated from the following relation:

$$\frac{\Sigma F_p}{\eta f_p^{max}} = N_P p \quad (4)$$

Where $\Sigma F_P$ is the aggregated pinning force density, $f_p^{max}$ is the maximum normalized elementary pinning force ($f_p$), and $\eta$ is an efficiency factor. The $\eta$ value is 1 in the case of a plastic lattice, and the $\eta$ value is otherwise $f_p^{max}/B$, where $B$ is the bulk modulus of the material [45]. We can assume that $\eta = 1$, as pressure can shrink lattice parameters. The inset of Fig.7 shows the $N_p$ versus temperature plot at P=0 and P=1GPa. The $N_p$ values are found to be much greater at 14 K at P=1GPa as compared to $N_p$ at $P=0$GPa (nearly six times as great). It is well known that hydrostatic pressure induces pinning centres which, in turn, leads to huge values of $J_c$ and increase in $N_p$ at P=1Gpa is a direct evidence of that [23, 46-48]. This is further verified in Fig.8, which shows the plot of $J_c/J_c^{max}$ versus reduced field (i.e. $h = H/H_{irr}$) at $P=0$GPa and $P=1$GPa for 14K and $P=0$GPa & $P=0.45$GPa for 12K. Obviously, the hump or secondary peak effect observed at high pressures suggests that the $J_c$ enhancement is due to induced pinning centres.

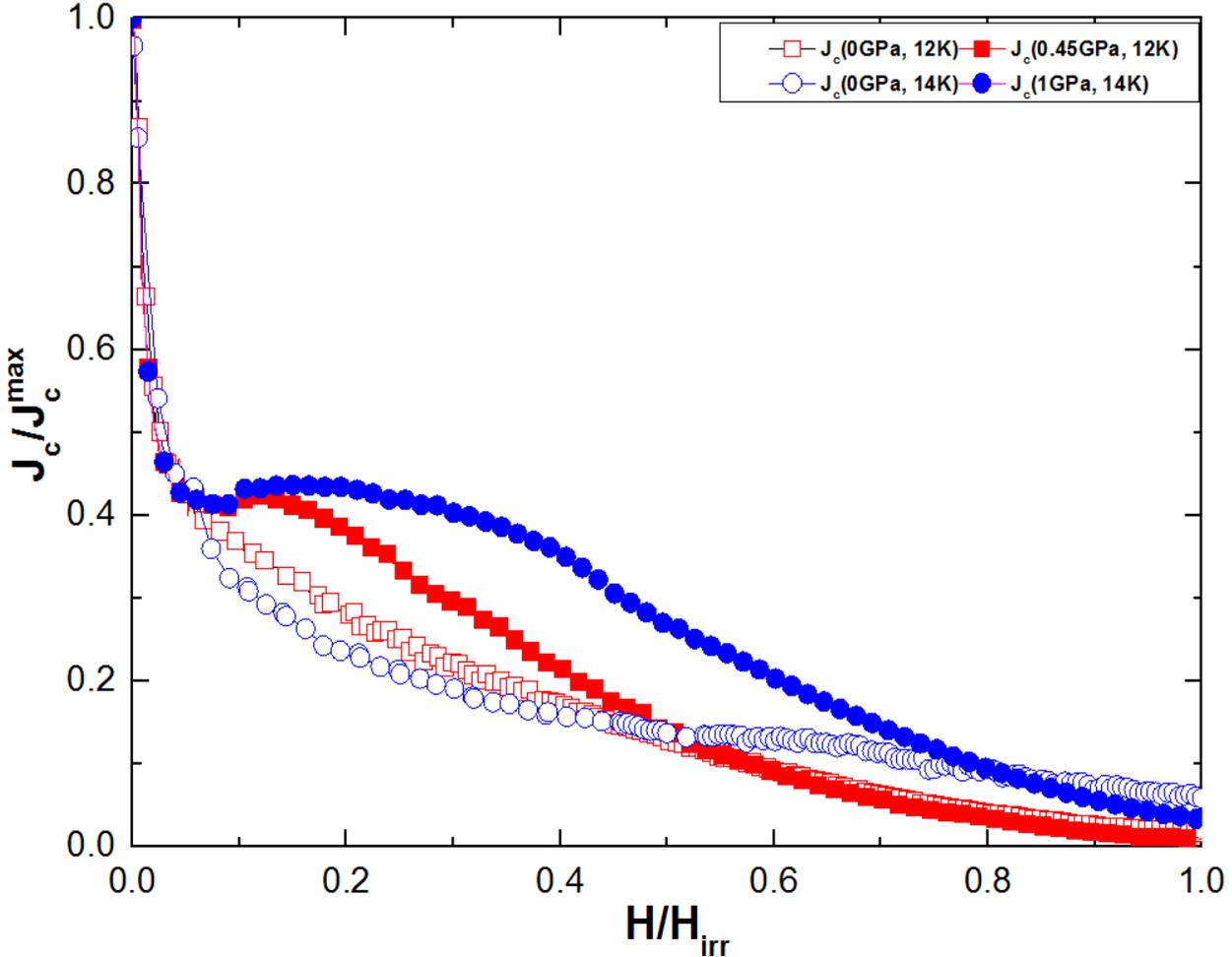

Figure 8: Reduced field dependence of the normalized $J_c$ at 14K for different pressures. The inset shows the same plot for 12K at $P=0$GPa and $P=0.45$Gpa.

Additionally, we found a pronounced reduction in the superconducting anisotropy at high temperatures, by almost 63% at $P=1$GPa. The pressure dependence of the $T_c$, volume ($V$) and anisotropy ($\gamma$) are interconnected through a relation [49]:

$$-\left[\frac{T_c^P - T_c^0}{T_c^0}\right] = \left[\frac{\Delta V}{V}\right] + \left[\frac{\gamma_P - \gamma_0}{\gamma_0}\right] \quad (5)$$

At $\Delta P$=1GPa, $\Delta V(T_c)/V(T_c)$ is estimated to be -0.02, as $\Delta V/V=-\Delta P/B$, where $K$ is the bulk modulus. We can use the bulk modulus (B≈52.3GPa) of a similar superconductor, i.e. Na$_{1-x}$FeAs [34]. The value of $\gamma$ at $P$=1GPa is found to be 63% less than its value at $P$=0GPa.

Hydrostatic pressure can significantly enhance $J_c$ by up to $10^2$ times in NaFe$_{0.97}$Co$_{0.03}$As single crystals, which is a record high enhancement. The most significant enhancement in in-field performance of NaFe$_{0.97}$Co$_{0.03}$As in terms of pinning force density ($F_p$) is found at $P$=1GPa in particular, where the $F_p$ at high fields is over 20 and 80 times higher at 12 and 14K, respectively, than at 0 GPa. The hydrostatic pressure induces more effective point pinning centres and $N_p$ at 1GPa is almost two times higher at 12K and over six times higher at 14K compared to the value at 0GPa. Moreover, a hump or secondary peak effect is found from the plot of the normalized $J_c$ as a function of reduced field. Therefore, this giant enhancement in $J_c$ values for NaFe$_{0.97}$Co$_{0.03}$As exists because of more pinning centres induced by pressure and the increase in pinning strength as well. The present study indicates that the supercurrent carrying ability in the Fe111 can be further and significantly increased by the proposed hydrostatic pressure technique. Our results were achived in single crystal samples, which means that the enhancement is intrinsic, and more significant than other reported approaches. It gives us high expections that the tapes or wires made by the same compounds should carry higher supercurrents using the hydrostatic pressure than those at ambient pressure.

Experimental

High-quality single crystals of NaFe$_{0.97}$Co$_{0.03}$As have been grown by use of the NaAs flux method. NaAs was obtained by reacting the mixture of the elemental Na and As in anevacuated quartz tube at 200°C for 10 h. Then NaAs, Fe,and Co powders were carefully weighed according to the ratio of NaAs:Fe:Co = 4:0.972:0.028, and thoroughly ground. The mixtures were put into alumina crucibles and then sealed in iron crucibles under 1.5 atm of highly pure argon gas. The sealed crucibles were heated to 950°C at a rate of 60°C/h in the tube furnace filled with the inert atmosphere and kept at 950°C for 10 h and then cooled slowly to 600°C at 3°C/h to grow single crystals.

The temperature dependence of the magnetic moments and the M-H loops at different temperatures and pressures were performed on Quantum Design Physical Property Measurement System (QD PPMS 14T) by using Vibrating Sample Magnetometer (VSM). We have used HMD High Pressure cell and Daphne 7373 oil as a pressure transmitting medium to apply hydrostatic pressure on a sample. The critical current density was calculated by using the Bean approximation.

Acknowledgments

X.L.W. acknowledges the support from the Australian Research Council (ARC) through an ARC Discovery Project (DP130102956) and an ARC Professorial Future Fellowship project (FT130100778).

Author Contributions

X.L.W conceived the pressure effects and designed the experiments. B.S. performed high pressure measurements. X.H.C. and A.F.W provided samples. X.L.W and B.S analysed the data and wrote the paper. S.R.G and S.D contributed to the discussions of the data and the paper.

Additional information

Competing financial interests: The authors declare no competing financial interests.